\documentclass{jpsj-suppl}
\usepackage{graphicx}
\usepackage{epsf}
\def\lamb#1#2{$^{#1}_{\Lambda}${#2}}
\def\lam#1#2{$^{#1}_{~\Lambda}${#2}}

\title{Charge symmetry breaking in $\Lambda$ hypernuclei: \\ 
updated HYP 2015 progress report} 
\author{Avraham \textsc{Gal}} 
\inst{Racah Institute of Physics, The Hebrew University, Jerusalem 91904, 
ISRAEL} 

\abst{Ongoing progress in understanding and evaluating charge symmetry 
breaking in $\Lambda$ hypernuclei is discussed in connection to recent 
measurements of the \lamb{4}{H}($0^+_{\rm g.s.}$) binding energy at 
MAMI [A1 Collaboration: PRL \textbf{114} (2015) 232501] and of 
the \lamb{4}{He}($1^+_{\rm exc}$) excitation energy at J-PARC 
[E13 Collaboration: PRL \textbf{115} (2015) 222501].} 

\kword{charge symmetry breaking, hypernuclei, hyperon-nucleon interaction 
models} 

\begin{document}
\maketitle

\section{Introduction} 

Charge symmetry in hadronic physics is broken in QCD by the light $u$--$d$ 
quark mass difference and by their QED interactions, both of which contribute 
significantly to the observed 1.3~MeV $n$--$p$ mass difference. In nuclear 
physics, charge symmetry breaking (CSB) results in a difference between the 
$nn$ and $pp$ scattering lengths, and also contributes about 70 keV out of 
the Coulomb-dominated 764~keV binding-energy difference in the mirror nuclei 
$^3$H and $^3$He, as reviewed in Ref.~\cite{miller06}. It can be explained by 
$\rho^0\omega$ mixing in one-boson exchange models of the $NN$ interaction, 
or by considering $N\Delta$ intermediate-state mass differences in models 
limited to pseudoscalar meson exchanges \cite{mach01}. 
In practice, introducing two charge dependent contact interaction terms 
in chiral effective field theory ($\chi$EFT) applications, one is able at 
next-to-next-to-next-to-leading order (N$^3$LO) to account quantitatively 
for the charge dependence of the low energy nucleon-nucleon ($NN$) scattering 
parameters and, thereby, also for the $A$=3 mirror nuclei binding-energy 
difference~\cite{entem03}. 

\begin{figure}[hbt] 
\begin{center} 
\includegraphics[scale=0.25]{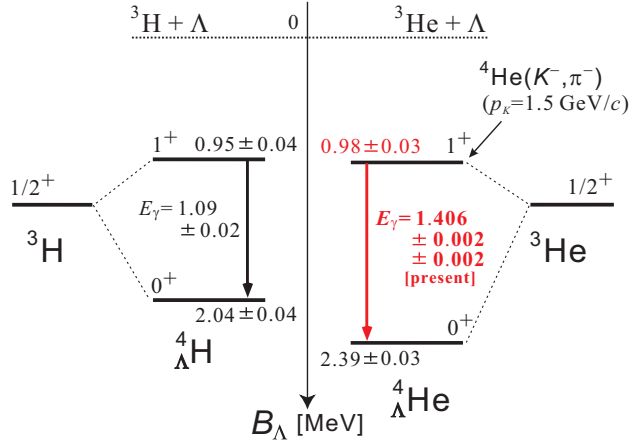} 
\caption{\lamb{4}{H}--\lamb{4}{He} level diagram. Ground-state separation 
energies $B_{\Lambda}$, loosely termed $\Lambda$ binding energies, are 
from emulsion data \cite{davis05}, and the $1^{+}_{\rm exc}$ excitation 
energies from $\gamma$-ray measurements \cite{E13}.} 
\label{fig:A=4} 
\end{center}
\end{figure}

In $\Lambda$ hypernuclei, with scarce and imprecise $\Lambda p$ scattering 
data and no $\Lambda n$ data, the only clear CSB signal is the large $\Lambda$ 
separation-energy difference $\Delta B^{J=0}_{\Lambda}$=350$\pm$60~keV in 
the $A$=4 $0^+_{\rm g.s.}$ hypernuclear mirror levels from old emulsion data 
\cite{davis05}, in contrast to the small difference $\Delta B^{J=1}_{\Lambda}$ 
in the $1^+_{\rm exc}$ states \cite{E13}, as shown in Fig.~\ref{fig:A=4}. 
Recent measurements \cite{MAMI15,MAMI16} at the Mainz Microtron (MAMI) of 
the $_{\Lambda}^4{\rm H}_{\rm g.s.}\to{^4{\rm He}}+\pi^-$ decay have produced 
a value of $B_{\Lambda}({_{\Lambda}^4{\rm H}_{\rm g.s.}})$=2.157$\pm$0.077~MeV 
\cite{MAMI16}, thereby confirming a substantial CSB $0^+_{\rm g.s.}$ splitting 
$\Delta B^{J=0}_{\Lambda}$=233$\pm$92~keV. This hypernuclear CSB ground state 
(g.s.) splitting is much larger than the $\approx$70~keV or so assigned to CSB 
splitting in the mirror core nuclei $^3$H and $^3$He. 

This updated CSB review, starting with work reported in Ref.~\cite{gal15}, 
demonstrates that the observed CSB splitting of mirror levels in the $A$=4 
$\Lambda$ hypernuclei can be reproduced by incorporating $\Lambda-\Sigma^0$ 
mixing \cite{DvH64} within a schematic $\Lambda N\leftrightarrow\Sigma N$ 
($\Lambda\Sigma$) coupling potential model for $s$-shell $\Lambda$ hypernuclei 
\cite{akaishi00,akaishi02}. It is further shown, by extending this schematic 
model to the $p$ shell \cite{millener08}, that smaller and perhaps negative 
CSB splittings result in mirror $p$-shell $\Lambda$ hypernuclear g.s. 
\cite{gal15}, in agreement with emulsion data \cite{davis05}. Finally, new 
results are presented from application of the J\"{u}lich-Bonn leading-order 
$\chi$EFT $YN$ interaction model \cite{LO06} in a complete four-body no-core 
shell model (NCSM) calculaion of the $A$=4 $\Lambda$ hypernuclei, again 
demonstrating that large CSB splittings can be obtained \cite{gg15,gg16}.

\section{CSB from $\Lambda-\Sigma^0$ mixing} 

Pion emission or absorption by a $\Lambda$ hyperon is forbidden by isospin, 
hence there is no one-pion exchange (OPE) contribution to the $\Lambda N$ 
charge symmetric (CS) strong interaction. However, as pointed out by 
Dalitz and von Hippel \cite{DvH64} the SU(3) octet $\Lambda_{I=0}$ and 
$\Sigma^0_{I=1}$ hyperons are admixed in the physical $\Lambda$ hyperon, 
thereby generating a direct $\Lambda N$ CSB potential $V_{\rm CSB}$ with 
a long-range OPE component that contributes substantially to the $0^+_{
\rm g.s.}$ splitting $\Delta B^{J=0}_{\Lambda}$ in the $A$=4 mirror 
hypernuclei. With updated coupling constants, their $0^+_{\rm g.s.}$ purely 
central wavefunction yields $\Delta B_{\Lambda}^{\rm OPE}(0^+_{\rm g.s.})
\approx 95$~keV. This is confirmed in our recent calculations in which tensor 
contributions add roughly another 100 keV \cite{gg16}. Shorter-range CSB 
meson-mixing contributions apparently are considerably smaller \cite{coon99}. 

The $\Lambda-\Sigma^0$ mixing mechanism gives rise also to a variety of 
(e.g. $\rho$) meson exchanges other than OPE. In baryon-baryon models 
that include {\it explicitly} a CS strong-interaction $\Lambda\Sigma$ 
coupling, the direct $\Lambda N$ matrix element of $V_{\rm CSB}$ is 
related to a suitably chosen strong-interaction isospin $I_{NY}=1/2$ 
matrix element $\langle N\Sigma|V_{\rm CS}|N\Lambda\rangle$ by 
\begin{equation} 
\langle N\Lambda|V_{\rm CSB}|N\Lambda\rangle = -0.0297\,\tau_{Nz}\,\frac{1}
{\sqrt{3}}\,\langle N\Sigma|V_{\rm CS}|N\Lambda\rangle , 
\label{eq:OME} 
\end{equation}  
where the isospin Clebsch-Gordan coefficient $1/\sqrt{3}$ accounts for the 
$N\Sigma^0$ amplitude in the $I_{NY}=1/2$ $N\Sigma$ state, and the space-spin 
structure of this $N\Sigma$ state is taken identical with that of the 
$N\Lambda$ state sandwiching $V_{\rm CSB}$. The $\approx$3\% CSB scale 
factor $-$0.0297 in (\ref{eq:OME}) follows from the matrix element of the 
$\Lambda-\Sigma^0$ mass mixing operator $\delta M$, 
\begin{equation} 
-2\;\frac{\langle\Sigma^0|\delta M|\Lambda\rangle}
{M_{\Sigma^0}-M_{\Lambda}}=-0.0297, 
\label{eq:DvH} 
\end{equation} 
by using for $\delta M$ one of the SU(3) mass formulae \cite{DvH64,galprd15} 
\begin{equation} 
\langle\Sigma^0|\delta M|\Lambda\rangle=\frac{1}{\sqrt 3}
(M_{\Sigma^0}-M_{\Sigma^+}+M_p-M_n)=1.14\pm 0.05~{\rm MeV}.   
\label{eq:deltaM} 
\end{equation} 
Lattice QCD calculations yield so far only half of this value for the 
mass-mixing matrix element \cite{lqcd15}. The reason apparently is 
the omission of QED from these calculations. 

\begin{figure}[t!] 
\begin{center} 
\includegraphics[scale=0.5]{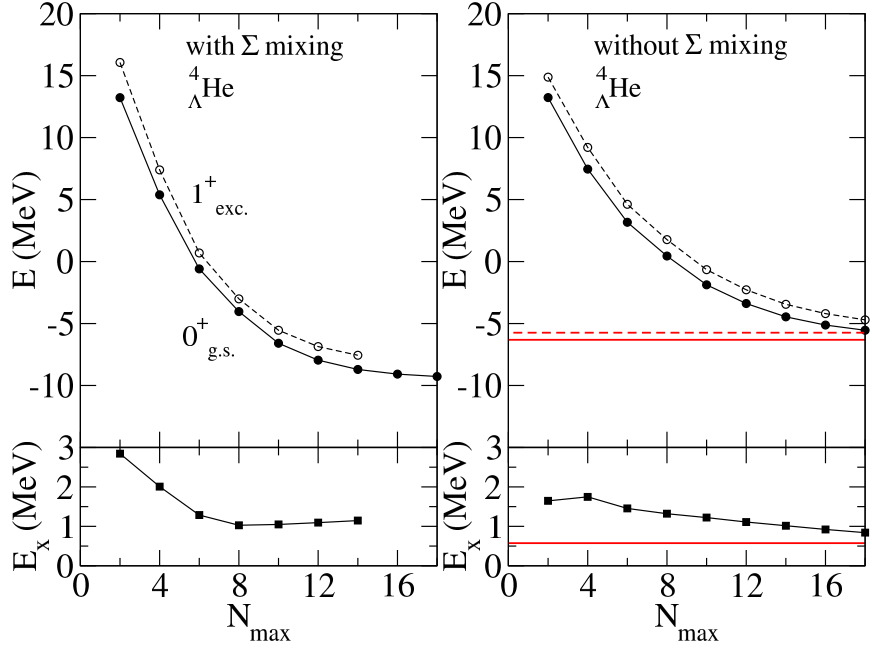} 
\caption{Energy eigenvalues $E$ and excitation energies $E_x$ in NCSM 
calculations of $_{\Lambda}^4{\rm He}(0^+_{\rm g.s.},1^+_{\rm exc})$ states 
\cite{gazda14,wirth14} as a function of $N_{\rm max}$, using LO $\chi$EFT $YN$ 
interactions with cutoff 600~MeV \cite{LO06}, including (left) or excluding 
(right) $\Lambda\Sigma$ coupling.} 
\label{fig:LO} 
\end{center} 
\end{figure} 

Since the CS strong-interaction $\Lambda\Sigma$ coupling, according to 
Eq.~(\ref{eq:OME}), is the chief provider of the CSB $\Lambda N$ matrix 
element, it is natural to ask how strong the $\Lambda\Sigma$ coupling is in 
realistic microscopic $YN$ interaction models. In Fig.~\ref{fig:LO} we show 
results of NCSM calculations of $_{\Lambda}^4{\rm He}$ levels \cite{gazda14}, 
using the J\"{u}lich-Bonn LO $\chi$EFT $YN$ CS potential model \cite{LO06},  
in which $\Lambda\Sigma$ coupling is seen to contribute almost 40\% of the 
$0^+_{\rm g.s.}\to 1^+_{\rm exc}$ excitation energy $E_x$. This also occurs 
in the Nijmegen NSC97 models \cite{NSC97} as demonstrated in the next section. 
With $\Lambda\Sigma$ matrix elements of order 10~MeV, the 3\% CSB scale factor 
(\ref{eq:DvH}) suggests CSB splittings of order 300~keV, in agreement with the 
observed $0^+_{\rm g.s.}$ CSB splitting, see Fig.~\ref{fig:A=4}.

\section{CSB in $s$-shell hypernuclei}

Akaishi et al.~\cite{akaishi00} derived $G$-matrix $YN$ effective interactions 
from NSC97 models \cite{NSC97}. These have been employed in Ref.~\cite{gal15} 
to calculate CSB contributions using Eq.~(\ref{eq:OME}) in which 
a spin-dependent central CS form is assumed for the $\Lambda\Sigma$ 
$0s_N0s_{Y}$ effective interaction $V_{\Lambda\Sigma}$, 
\begin{equation} 
V_{\Lambda\Sigma}=({\bar V}_{\Lambda\Sigma}+\Delta_{\Lambda\Sigma}{\vec s}_N
\cdot{\vec s}_{Y})\sqrt{4/3}\;{\vec t}_N\cdot{\vec t}_{\Lambda\Sigma}, 
\label{eq:V_YN} 
\end{equation} 
and where ${\vec t}_{\Lambda\Sigma}$ converts a $\Lambda$ to $\Sigma$ in 
isospace. The $s$-shell $0s_N0s_Y$ matrix elements ${\bar V}^{0s}_{\Lambda
\Sigma}$ and $\Delta^{0s}_{\Lambda\Sigma}$ are listed in Table~\ref{tab:A=4}, 
adapted from Ref.~\cite{gal15}, for two such $G$-matrix models denoted 
$(\Lambda\Sigma)_{\rm e,f}$. The $A$=4 matrix elements $v(J^{\pi})$, in terms 
of these two-body matrix elements, are 
\begin{equation} 
v(0^+_{\rm g.s.})={\bar V}^{0s}_{\Lambda\Sigma}+\frac{3}{4}
\Delta^{0s}_{\Lambda\Sigma},\;\;\; v(1^+_{\rm exc})={\bar V}^{0s}_{\Lambda
\Sigma}-\frac{1}{4}\Delta^{0s}_{\Lambda\Sigma},  
\label{eq:v(JP)} 
\end{equation} 
from which the downward energy shifts $\delta E_{\downarrow}(J^{\pi})$ defined 
by $\delta E_{\downarrow}(J^{\pi})= v^2(J^{\pi})/(80~{\rm MeV})$ are readily 
evaluated, with their difference $E_x^{\Lambda\Sigma}$ listed in the 
table. Furthermore, by comparing this partial excitation-energy contribution 
to the listed values of the total $E_x(0^+_{\rm g.s.} - 1^+_{\rm exc})$ from 
Refs.~\cite{akaishi00,akaishi02,nogga02} we demonstrate a sizable $\sim$50\% 
contribution of $\Lambda\Sigma$ coupling to the observed excitation energy 
$E_x(0^+_{\rm g.s.} - 1^+_{\rm exc})\approx 1.25$~MeV deduced from the 
$\gamma$-ray transition energies marked in Fig.~\ref{fig:A=4}. Recall also 
the sizable $\Lambda\Sigma$ contribution to $E_x$ shown in Fig.~\ref{fig:LO} 
for the NCSM calculation \cite{gazda14} using the J\"{u}lich-Bonn LO $\chi$EFT 
$YN$ interaction model \cite{LO06}.  

\begin{table}[htb]
\begin{center}
\caption{$\Lambda\Sigma$ $s$-shell matrix elements 
${\bar V}^{0s}_{\Lambda\Sigma}$ and $\Delta^{0s}_{\Lambda\Sigma}$ in 
models $(\Lambda\Sigma)_{\rm e,f}$ \cite{millener08} and the resulting 
$\Lambda\Sigma$ contribution $E_x^{\Lambda\Sigma}$ to the $0^+_{\rm g.s.}\to 
1^+_{\rm exc}$ excitation energy in the $A=4$ hypernuclear states. The total 
excitation energy $E_x(0^+_{\rm g.s.}-1^+_{\rm exc})$ and CSB splittings 
$\Delta B_{\Lambda}(J^{\pi})$ calculated in several models are also given. 
Note that $\Delta B_{\Lambda}(J^{\pi})=0.0343\,v(J^{\pi})$ in the schematic 
model \cite{gal15}. Listed values are in MeV.} 
\begin{tabular}{ccccccccccc} 
\hline \hline 
NSC97 & ${\bar V}^{0s}_{\Lambda\Sigma}$ & $\Delta^{0s}_{\Lambda\Sigma}$ & 
$E_x^{\Lambda\Sigma}$ & \multicolumn{3}{c}{$E_x(0^+_{\rm g.s.}-1^+_{
\rm exc})$} & \multicolumn{2}{c}{$\Delta B_{\Lambda}(0^+_{
\rm g.s.})$} & \multicolumn{2}{c}{$\Delta B_{\Lambda}(1^+_{\rm exc})$}  \\ 
\cite{NSC97} & \multicolumn{3}{c}{$(\Lambda\Sigma)_{\rm e,f}$ models 
\cite{millener08}} & \cite{akaishi00} & \cite{akaishi02} & \cite{nogga02} & 
\cite{nogga02,nogga07} & \cite{gal15} & \cite{nogga02,nogga07} & \cite{gal15} 
\\ 
\hline 
NSC97$_{\rm e}$ & 2.96 & 5.09 & 0.539 & 0.89 & 1.13 & 0.79 & 0.075 & 0.226 & 
$-$0.010 & 0.030 \\ 
NSC97$_{\rm f}$ & 3.35 & 5.76 & 0.689 & 1.48 & 1.51 & 1.16 & 0.100 & 0.266 & 
$-$0.010 & 0.039 \\ 
\hline \hline 
\end{tabular} 
\label{tab:A=4} 
\end{center} 
\end{table} 

Listed in the last four columns of Table~\ref{tab:A=4} are $A$=4 CSB 
splittings $\Delta B_{\Lambda}(J^{\pi})$, calculated for NSC97 $YN$ models in 
Refs.~\cite{nogga02,nogga07} and for the schematic $\Lambda\Sigma$ coupling 
model in Ref.~\cite{gal15}. The listed CSB splittings include a residual 
($V_{\rm CSB}=0$) splitting of size $\approx$30~keV consisting of a small 
positive contribution from the $\Sigma^{\pm}$ mass difference and a small 
negative contribution from the slightly increased Coulomb repulsion in 
$_{\Lambda}^4{\rm He}$ with respect to that in its $^3$He core. 
The $1^+_{\rm exc}$ CSB splittings listed in the table come out universally 
small in these models owing to the specific spin dependence of $V_{\Lambda
\Sigma}$. The values of $\Delta B_{\Lambda}(0^+_{\rm g.s.})$ listed in 
Table~\ref{tab:A=4} are smaller than 100~keV upon using NSC97 models, thereby 
leaving the $A=4$ CSB puzzle unresolved, while being larger than 200~keV in 
the schematic $\Lambda\Sigma$ model and therefore getting considerably closer 
to the experimentally reported $0^+_{\rm g.s.}$ CSB splitting. A direct 
comparison between the NCS97 models and the schematic $\Lambda\Sigma$ model 
is not straightforward because the $\Lambda\Sigma$ coupling in NSC97 models 
is dominated by tensor components, whereas no tensor components appear in the 
schematic $\Lambda\Sigma$ model. 

\begin{figure}[hbt]
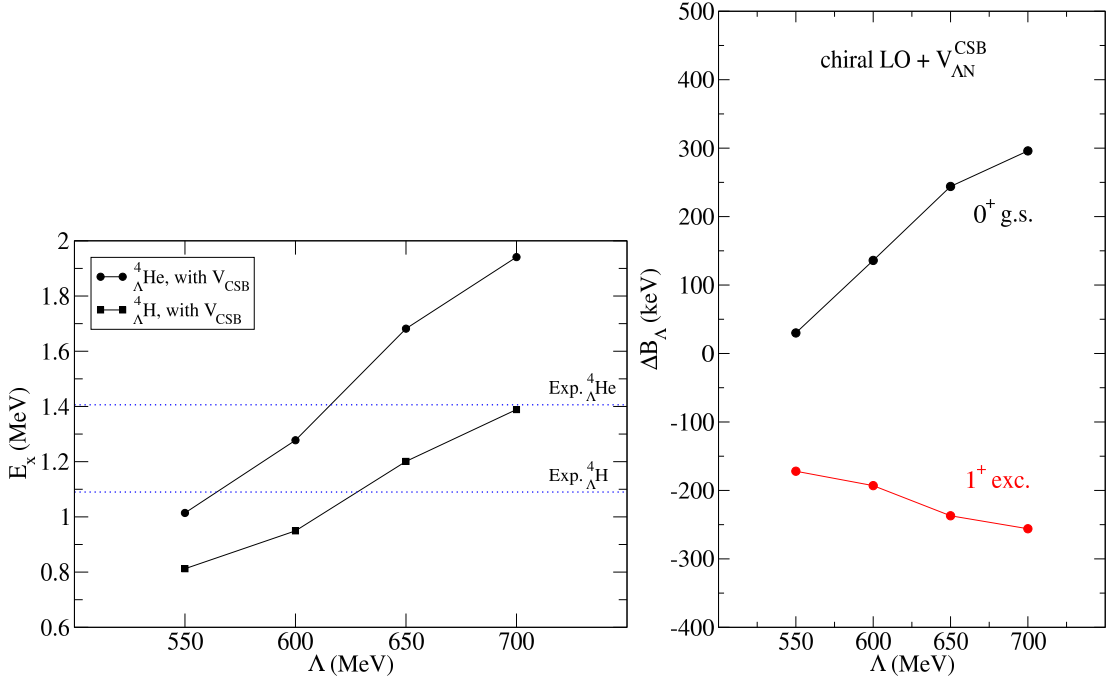
 
\begin{center} 
\includegraphics[scale=0.35,clip]{ex-l_vm_csb.eps} 
\includegraphics[scale=0.35,clip]{ldep_csbls_rev.eps} 
\caption{Cutoff momentum dependence of excitation energies 
$E_{\rm x}$(0$^+_{\rm g.s.}$$\to$1$^+_{\rm exc}$) (left) and of CSB splittings 
$\Delta B_{\Lambda}(J^{\pi})$ (right) in NCSM calculations \cite{gg15,gg16} 
of the $A$=4 hypernuclei, using LO $\chi$EFT $YN$ interactions \cite{LO06}. 
Values of $E_{\rm x}$ from $\gamma$-ray measurements \cite{E13} are marked by 
dotted horizontal lines.}  
\label{fig:gg15} 
\end{center} 
\end{figure} 

Results of recent four-body NCSM calculations of the $A$=4 hypernuclei 
\cite{gg15,gg16}, using the Bonn-J\"{u}lich LO $\chi$EFT SU(3)-based 
$YN$ interaction model \cite{LO06} with cutoff momentum in the range 
$\Lambda$=550--700~MeV, are shown in Fig.~\ref{fig:gg15}. In line with the 
schematic model, the $\Lambda\Sigma$ coupling potential in this $\chi$EFT 
model is dominated by a central-interaction contact term. Plotted on the 
left-hand side (l.h.s.) are the calculated $0^+_{\rm g.s.}\to 1^{+}_{\rm exc}$ 
excitation energies $E_{\rm x}$, for which the CS $\Lambda\Sigma$ coupling 
potential according to Fig.~\ref{fig:LO} is so crucial. With $\Lambda$ between 
600 and 650 MeV, one is close to reproducing the $\gamma$-ray measured values 
of $E_{\rm x}$. In fact for $\Lambda$=600~MeV the induced CSB generates 
a value of $\Delta B_{\Lambda}^{\rm calc}(0^+_{\rm g.s.})-\Delta B_{
\Lambda}^{\rm calc}(1^+_{\rm exc})=330\pm 40$~keV, in excellent agreement with 
the measured value of $E_{\rm x}$(\lamb{4}{He})$-E_{\rm x}$(\lamb{4}{H})$=320 
\pm 20$~keV, see Fig.~\ref{fig:A=4}. Other models underestimate this measured 
value of $\Delta E_{\rm x}$, with $\approx$210~keV in the schematic $\Lambda
\Sigma$ model and at most $\approx$110~keV in the NSC97$_{\rm f}$ model. 
Plotted on the right-hand side of Fig.~\ref{fig:gg15} are the separate CSB 
splittings $\Delta B_{\Lambda}(J^{\pi})$, demonstrating for the first time 
that the observed CSB splitting of the $0^+_{\rm g.s.}$ mirror levels can be 
reproduced using realistic theoretical interaction models, although with 
appreciable momentum cutoff dependence. We note that the central value of 
$\Delta B_{\Lambda}^{\rm exp}(0^+_{\rm g.s.})$=233$\pm$92~keV, as derived 
from the recent measurement of $B_{\Lambda}({_{\Lambda}^4{\rm H}})$ at 
MAMI~\cite{MAMI16}, is comfortably reproduced for $\Lambda$=650~MeV.

\section{CSB in $p$-shell hypernuclei}

Recent work by Hiyama {\it et al.} has failed to explain CSB splittings in 
$p$-shell mirror hypernuclei \cite{hiyama09,zhang12,hiyama12}, apparently for 
disregarding the underlying CS $\Lambda\Sigma$ coupling potential. In the 
approach reviewed here, one extends the NSC97$_{\rm e}$ model $0s_N0s_{Y}$ 
effective interactions by providing $(\Lambda\Sigma)_{\rm e}$ $0p_N0s_Y$ 
central-interaction matrix elements which are consistent with the role 
$\Lambda\Sigma$ coupling plays in a shell-model reproduction of hypernuclear 
$\gamma$-ray transition energies by Millener \cite{millener12}. The $p$-shell 
$0p_N0s_Y$ matrix elements (given in the caption to Table~\ref{tab:pshell}) 
are smaller by roughly a factor of two from the $s$-shell $0s_N0s_{Y}$ matrix 
elements in Table~\ref{tab:A=4}, reflecting the reduced weight which the 
major relative $s$-wave matrix elements of $V_{NY}$ assume in the $p$ shell. 
This suggests that $\Sigma$ admixtures, which are quadratic in these matrix 
elements, are weaker roughly by a factor of four with respect to the $s$-shell 
calculation, and also that CSB contributions in the $p$ shell are weaker 
with respect to those in the $A=4$ hypernuclei, although only by a factor 
of two. To evaluate these CSB contributions, the single-nucleon expression 
(\ref{eq:OME}) is extended by summing over the $p$-shell nucleons:  
\begin{equation} 
V_{\rm CSB} = -0.0297\,\frac{1}{\sqrt{3}}
\sum_j{({\bar V}^{0p}_{\Lambda\Sigma}+\Delta^{0p}_{\Lambda\Sigma}
{\vec s}_j\cdot{\vec s}_Y)\tau_{jz}}. 
\label{eq:VCSB} 
\end{equation} 

Results of applying the present $(\Lambda\Sigma)_{\rm e}$ coupling model to 
several pairs of g.s. levels in $p$-shell hypernuclear isomultiplets are given 
in Table~\ref{tab:pshell}, extended from Ref.~\cite{gal15}. All pairs except 
for $A=7$ are mirror hypernuclei identified in emulsion \cite{davis05} where 
binding energy systematic uncertainties are largely canceled out in forming 
the listed $\Delta B_{\Lambda}^{\rm exp}$ values. For $A=7$ we calculated 
(i) $\Delta B_{\Lambda}$(\lamb{7}{Be}$-$\lamb{7}{Li}$^{\ast}$), comparing 
it to $\Delta B_{\Lambda}$ obtained from g.s. emulsion data, as well as 
(ii) $\Delta B_{\Lambda}$(\lamb{7}{Li}$^{\ast}-$\lamb{7}{He}), comparing 
it to $\Delta B_{\Lambda}$ obtained from FINUDA $\pi^-$-decay data 
for \lamb{7}{Li}$_{\rm g.s.}$~\cite{botta09} and from very recent JLab 
electroproduction data for \lamb{7}{He}~\cite{JLabL7He}. The Jlab and 
FINUDA measurements allow comparison since by using magnetic spectrometers 
it becomes possible to make absolute energy calibrations relative to precise 
values of free-space known masses. Note that the value reported by FINUDA for 
$B_{\Lambda}$(\lamb{7}{Li}$_{\rm g.s.}$), 5.85$\pm$0.17~MeV, differs from 
the emulsion value of 5.58$\pm$0.05~MeV (including systematic errors too, 
see \cite{botta16}). To obtain $B_{\Lambda}$(\lamb{7}{Li}$^{\ast}$) 
from $B_{\Lambda}$(\lamb{7}{Li}$_{\rm g.s.}$) we made use 
of the observation of the 3.88~MeV $\gamma$-ray transition 
\lamb{7}{Li}$^{\ast}\to\gamma$+\lamb{7}{Li}~\cite{tamura00}. Note that the 
$^6$Li core state of \lamb{7}{Li}$^{\ast}$ is the $0^+$ $T$=1 at 3.56 MeV, 
whereas the core state of \lamb{7}{Li}$_{\rm g.s.}$ is the $1^+$ $T$=0 g.s. 
Recent $B_{\Lambda}$ values from JLab electroproduction experiments at JLab 
for \lamb{9}{Li}~\cite{JLabL9Li} and \lam{10}{Be}~\cite{JLabL10Be} were not 
used for lack of similar data on their mirror partners. 

\begin{table}[htb]
\begin{center}
\caption{CSB contributions to $\Delta B^{\rm calc}_{\Lambda}(\rm g.s.)$ values 
in $p$-shell hypernuclear isomultiplets, using the $(\Lambda\Sigma)_{\rm e}$ 
coupling model with matrix elements ${\bar V}^{0p}_{\Lambda\Sigma}=1.45$ 
and $\Delta^{0p}_{\Lambda\Sigma}=3.04$~MeV in Eq.~(\ref{eq:VCSB}); see text. 
The $s$-shell contributions to $\Delta B_{\Lambda}(0^+_{\rm g.s.})$ from 
Table~\ref{tab:A=4} are also listed for comparison. Listed values of 
$\Delta B_{\Lambda}^{\rm exp}$ are based on g.s. emulsion data except for 
$\Delta B_{\Lambda}^{\rm exp}$(\lamb{7}{Li}$^{\ast}-$\lamb{7}{He}), see text.} 
\begin{tabular}{cclccrrr} 
\hline \hline 
\lamb{A}{Z$_{>}$}--\lamb{A}{Z$_{<}$} & $I,J^{\pi}$ & 
$P_{\Sigma}$ & $\Delta T_{YN}$ & $\Delta V_C$ & $\langle V_{\rm CSB}\rangle$ 
& $\Delta B_{\Lambda}^{\rm calc}$ & $\Delta B_{\Lambda}^{\rm exp}$ \\     
pairs & & (\%) & (keV) & (keV) & (keV) & (keV) & (keV) \\ 
\hline 
\lamb{4}{He}--\lamb{4}{H} & $\frac{1}{2},0^+$ & 0.72 & 39 & $-$45 & 232 & 
226 & $+$350$\pm$60 \\
\hline 
\lamb{7}{Be}--\lamb{7}{Li}$^{\ast}$ & $1,{\frac{1}{2}}^+$ & 0.12 & 3 & 
$-$70 \cite{hiyama09} & 50 & $-$17 & $-$100$\pm$90 \\ 
\lamb{7}{Li}$^{\ast}$--\lamb{7}{He} & $1,{\frac{1}{2}}^+$ & 0.12 & 2 & 
$-$80 \cite{hiyama09} & 50 & $-$28 & $-$20$\pm$230 \\ 
\lamb{8}{Be}--\lamb{8}{Li} & $\frac{1}{2},1^-$ & 0.20 & 11 & 
$-$81 \cite{hiyama02a} & 119 & $+$49 & $+$40$\pm$60 \\ 
\lamb{9}{B}--\lamb{9}{Li} & $1,{\frac{3}{2}}^+$ & 0.23 & 10 & 
$-$145 & 81 & $-$54 & $-$210$\pm$220 \\ 
\lam{10}{B}--\lam{10}{Be} & $\frac{1}{2},1^-$ & 0.053 & 3 & 
$-$156 & 17 & $-$136 & $-$220$\pm$250 \\ 
\hline \hline 
\end{tabular} 
\label{tab:pshell} 
\end{center} 
\end{table}

The $\Sigma$ admixture probabilities $P_{\Sigma}$ in Table~\ref{tab:pshell} 
follow from $\Lambda\Sigma$ strong-interaction contributions to $p$-shell 
hypernuclear g.s. energies computed in Ref.~\cite{millener12}. The associated 
CSB kinetic-energy contributions $\Delta T_{YN}$ were calculated using values 
of $P_{\Sigma}$ and $\Sigma^{\pm}$ mass differences. These $\Delta T_{YN}$ 
contributions, of order 10~keV and less, are considerably weaker than those 
for $A$=4 in the $s$ shell, reflecting weaker $\Sigma$ admixtures in the $p$ 
shell as listed in the table. The Coulomb-induced contributions $\Delta V_C$ 
are dominated by their $\Delta V^{\Lambda}_C$ components which were taken from 
Hiyama's cluster-model calculations \cite{hiyama09,hiyama02a} for $A=7,8$ 
and from Millener's unpublished shell-model notes for $A=9,10$. These 
contributions are always negative owing to the increased Coulomb repulsion 
in the $\Lambda$ hypernucleus with respect to its core. The sizable negative 
$p$-shell $\Delta V_C$ contributions, in distinction from their secondary role 
in forming the total $s$-shell $\Delta B_{\Lambda}(0^+_{\rm g.s.})$, exceed 
in size the positive $p$-shell $\langle V_{\rm CSB}\rangle$ contributions by 
a large margin beginning with $A=9$, thereby resulting in clearly negative 
values of $\Delta B_{\Lambda}(\rm g.s.)$. 

The $\langle V_{\rm CSB}\rangle$ contributions listed in Table~\ref{tab:pshell} 
were calculated using weak-coupling $\Lambda$-hypernuclear shell-model 
wavefunctions in terms of the corresponding nuclear-core g.s. leading 
SU(4) supermultiplet components, except for $A=8$ where the first 
excited nuclear-core level had to be included. The listed $A=7-10$ 
values of $\langle V_{\rm CSB}\rangle$ exhibit strong SU(4) correlations, 
marked in particular by the enhanced value of 119~keV for the SU(4) 
nucleon-hole configuration in \lamb{8}{Be}--\lamb{8}{Li} with respect 
to the modest value of 17~keV for the SU(4) nucleon-particle configuration 
in \lam{10}{B}--\lam{10}{Be}. This enhancement follows from the relative 
magnitudes of the Fermi-like interaction term ${\bar V}^{0p}_{\Lambda\Sigma}$ 
and its Gamow-Teller partner term $\Delta^{0p}_{\Lambda\Sigma}$ listed 
in the caption to Table~\ref{tab:pshell}. Noting that both $A=4$ and 
$A=8$ mirror hypernuclei correspond to SU(4) nucleon-hole configuration, 
the roughly factor two ratio of $\langle V_{\rm CSB}{\rangle}_{A=4}$=232~keV 
to $\langle V_{\rm CSB}{\rangle}_{A=8}$=119~keV reflects the approximate 
factor of two discussed earlier for the ratio between $s$-shell to $p$-shell 
$\Lambda\Sigma$ matrix elements. 

Comparing $\Delta B_{\Lambda}^{\rm calc}$ with $\Delta B_{\Lambda}^{\rm exp}$ 
in Table~\ref{tab:pshell}, we note the reasonable agreement reached between 
the $(\Lambda\Sigma)_{\rm e}$ coupling model calculation and experiment for 
all five pairs of $p$-shell hypernuclei, $A=7-10$, listed here. Extrapolating 
to heavier hypernuclei, one might naively expect negative values of $\Delta 
B_{\Lambda}^{\rm calc}$. However, this rests on the assumption that the 
negative $\Delta V^{\Lambda}_C$ contribution remains as large upon increasing 
$A$ as it is in the beginning of the $p$ shell, which need not be the case. 
As nuclear cores beyond $A=9$ become more tightly bound, the $\Lambda$ 
hyperon is unlikely to compress these nuclear cores as much as it does 
in lighter hypernuclei, so that the additional Coulomb repulsion in 
\lam{12}{C}, for example, over that in \lam{12}{B}, while still negative, 
may not be sufficiently large to offset the attractive CSB contribution to 
$B_{\Lambda}$(\lam{12}{C})$-B_{\Lambda}$(\lam{12}{B}). Hence, one expects 
that $|\Delta B_{\Lambda}(A=12)| \lesssim 50$~keV, in agreement with the 
recent discussion of measured $B_{\Lambda}$ systematics \cite{botta16}. 
In making this argument one relies on the expectation, based on SU(4) 
supermultiplet fragmentation patterns in the $p$ shell, that $\langle 
V_{\rm CSB}\rangle$ does not exceed $\sim$100~keV. 

Some implications of the state dependence of CSB splittings, e.g. the large 
difference between the calculated $\Delta B_{\Lambda}(0^+_{\rm g.s.})$ and 
$\Delta B_{\Lambda}(1^+_{\rm exc})$ in the $s$ shell, are worth noting also 
in the $p$ shell, the most spectacular one concerns the \lam{10}{B} g.s. 
doublet splitting. Adding the $(\Lambda\Sigma)_{\rm e}$ coupling model CSB 
contribution of $\approx -27$~keV to the $\approx$110~keV CS $1^-_{\rm g.s.}
\to 2^-_{\rm exc}$ g.s. doublet excitation energy calculated in this model 
\cite{millener12} helps bring it down well below 100~keV, which is the upper 
limit placed on it from past searches for a $2^-_{\rm exc}\to 1^-_{\rm g.s.}$ 
$\gamma$-ray transition \cite{chrien90,tamura05}.

\section{Summary and outlook} 

The recent J-PARC observation of a 1.41 MeV \lamb{4}{He}($1^+_{\rm exc}\to 
0^+_{\rm g.s.}$) $\gamma$-ray transition \cite{E13}, and the recent MAMI 
determination of $B_{\Lambda}$(\lamb{4}{H}) to better than 100 keV 
\cite{MAMI15,MAMI16}, arose renewed interest in the sizable CSB confirmed 
thereby in the $A$=4 mirror hypernuclei. It was shown in the present updated 
report how a relatively large $\Delta B_{\Lambda}(0^+_{\rm g.s.})$ CSB 
contribution of order 250~keV arises in $\Lambda\Sigma$ coupling models based 
on Akaishi's $G$-matrix effective $s$-shell central interactions approach 
\cite{akaishi00,akaishi02}, well within the uncertainty of the value 
233$\pm$92~keV deduced from the recent MAMI measurement \cite{MAMI16}. It was 
also argued that the reason for the $YNNN$ coupled-channel calculations using 
NSC97 models to fall considerably behind, with 100~keV at most, is that their 
$\Lambda\Sigma$ coupling is dominated by a strong tensor term. In this sense, 
the observed large value of $\Delta B_{\Lambda}(0^+_{\rm g.s.})$ places 
a powerful constraint on the strong-interaction $YN$ dynamics. Recent results 
of ab-initio four-body calculations \cite{gg15,gg16} using $\chi$EFT $YN$ 
interactions in LO exhibit sizable CSB $0^+_{\rm g.s.}$ splittings in rough 
agreement with experiment. In future work one should apply the CSB generating 
equation (\ref{eq:OME}) in four-body calculations of the $A$=4 mirror 
hypernuclei using the available NLO $\chi$EFT version \cite{NLO13,nogga13}, 
and also to readjust the $\Lambda\Sigma$ contact terms in NLO by imposing the 
most accurate CSB datum as a further constraint. 

Finally, an extension of the schematic $\Lambda\Sigma$ coupling model to the 
$p$ shell was shown to reproduce successfully the main CSB features indicated 
by mirror-hypernuclei binding energies there \cite{gal15}. More theoretical 
work in this mass range, and beyond, is needed to understand further and 
better the salient features of $\Lambda\Sigma$ dynamics \cite{galmil13}. On 
the experimental side, the recenly approved J-PARC E63 experiment is scheduled 
to remeasure the \lamb{4}{H}($1^+_{\rm exc}\to 0^+_{\rm g.s.}$) $\gamma$-ray 
transition \cite{tamura16} and, perhaps in addition to the standard 
($\pi^+,K^+$) reaction, to also use the recently proposed ($\pi^-,K^0$) 
reaction \cite{agnello16} in order to study simultaneously several members 
of given $\Lambda$ hypernuclear isomultiplets, for example reaching both 
\lam{12}{B} and \lam{12}{C} on a carbon target.

\section*{Acknowledgments}

Fruitful collaboration with Daniel Gazda and shell-model guidance by John 
Millener are gratefully acknowledged, as well as the hospitality and support 
extended by Satoshi Nakamura and Hirokazu Tamura during HYP2015 at Sendai, 
Japan.


\begin{thebibliography}{99} 

\bibitem{miller06} G.A.~Miller, A.K.~Opper, and E.J.~Stephenson: 
Annu. Rev. Nucl. Part. Sci. \textbf{56} (2006) 253. 

\bibitem{mach01} R.~Machleidt and H.~M\"{u}ther: Phys. Rev. C \textbf{63} 
(2001) 034005. 

\bibitem{entem03} D.R.~Entem and R.~Machleidt: Phys. Rev. C \textbf{68} 
(2003) 041001(R). 

\bibitem{davis05} D.H.~Davis: Nucl. Phys. A \textbf{754} (2005) 3c. 

\bibitem{E13} T.O.~Yamamoto {\it et al.} (J-PARC E13 Collaboration): 
Phys. Rev. Lett. \textbf{115} (2015) 222501. 

\bibitem{MAMI15} A.~Esser {\it et al.} (MAMI A1 Collaboration): Phys. 
Rev. Lett. \textbf{114} (2015) 232501.   

\bibitem{MAMI16} F. Schulz {\it et al.} (MAMI A1 Collaboration): Nucl. 
Phys. A \textbf{954} (2016) 149.

\bibitem{gal15} A.~Gal: Phys. Lett. B \textbf{744} (2015) 352. 

\bibitem{DvH64} R.H.~Dalitz and F.~von Hippel: Phys. Lett. \textbf{10} 
(1964) 153. 

\bibitem{akaishi00} Y.~Akaishi, T.~Harada, S.~Shinmura, and K.S.~Myint: 
Phys. Rev. Lett. \textbf{84} (2000) 3539. 

\bibitem{akaishi02} H.~Nemura, Y.~Akaishi, and Y.~Suzuki: Phys. Rev. Lett. 
\textbf{89} (2002) 142504. 

\bibitem{millener08} D.J.~Millener: Nucl. Phys. A \textbf{754} (2005) 48c, 
{\it ibid.} \textbf{804} (2008) 84. 

\bibitem{LO06} H.~Polinder, J.~Haidenbauer, and U.-G.~Mei{\ss}ner: 
Nucl. Phys. A \textbf{779} (2006) 244. 

\bibitem{gg15} D.~Gazda and A.~Gal: Phys. Rev. Lett. \textbf{116} (2016) 
122501.  

\bibitem{gg16} D.~Gazda and A.~Gal: Nucl. Phys. A \textbf{954} (2016) 161. 

\bibitem{coon99} S.A.~Coon, H.K.~Han, J.~Carlson, and B.F.~Gibson: 
in {\it Meson and Light Nuclei '98}, edited by J.~Adam, P.~Byd\v{z}ovsk\'{y}, 
J.~Dobe\v{s}, R.~Mach, and J.~Mare\v{s} (WS, Singapore, 1999), pp.~407-413.  

\bibitem{galprd15} A.~Gal: Phys. Rev. D \textbf{92} (2015) 018501. 

\bibitem{lqcd15} R.~Horsley {\it et al.}: Phys. Rev. D \textbf{91} (2015) 
074512, {\it ibid.} \textbf{92} (2015) 018502. 

\bibitem{gazda14} D.~Gazda, J.~Mare\v{s}, P.~Navr\'{a}til, R.~Roth, and 
R.~Wirth: Few-Body Syst. \textbf{55} (2014) 857. 

\bibitem{wirth14} R.~Wirth, D.~Gazda, P.~Navr\'{a}til, A.~Calci, 
J.~Langhammer, and R.~Roth: Phys. Rev. Lett. \textbf{113} (2014) 192502. 

\bibitem{NSC97} Th.A.~Rijken, V.G.J.~Stoks, and Y.~Yamamoto: Phys. Rev. C 
\textbf{59} (1999) 21. 

\bibitem{nogga02} A.~Nogga, H.~Kamada, and W.~Gl\"{o}ckle: Phys. Rev. Lett. 
\textbf{88} (2002) 172501. 

\bibitem{nogga07} J.~Haidenbauer, U.-G.~Mei{\ss}ner, A.~Nogga, and 
H.~Polinder: in {\it Topics in Strangeness Nuclear Physics}, Lecture Notes 
in Physics \textbf{724}, edited by P.~Byd\v{z}ovsk\'{y}, J.~Mare\v{s}, 
and A. Gal (Springer, New York, 2007), pp.~113-140. 


\bibitem{hiyama09} E.~Hiyama, Y.~Yamamoto, T.~Motoba, and M.~Kamimura: 
Phys. Rev. C \textbf{80} (2009) 054321. 
 
\bibitem{zhang12} Y.~Zhang, E.~Hiyama, and Y.~Yamamoto: Nucl. Phys. A 
\textbf{881} (2012) 288.  

\bibitem{hiyama12} E.~Hiyama and Y.~Yamamoto: Prog. Theor. Phys. \textbf{128} 
(2012) 105. 

\bibitem{millener12} D.J.~Millener: Nucl. Phys. A \textbf{881} (2012) 298, 
and references listed therein. 

\bibitem{hiyama02a} E.~Hiyama, M.~Kamimura, T.~Motoba, T.~Yamada, and 
Y.~Yamamoto: Phys. Rev. C \textbf{66} (2002) 024007. 

\bibitem{botta09} M.~Agnello {\it et al.} (FINUDA Collaboration and A.~Gal): 
Phys. Lett. B \textbf{681} (2009) 139. 

\bibitem{JLabL7He} T.~Gogami {\it et al.} (JLab HKS Collaboration): Phys. 
Rev. C \textbf{94} (2016) 021302(R). 

\bibitem{botta16} E.~Botta, T.~Bressani, and A.~Feliciello: arXiv:1608.07448. 

\bibitem{tamura00} H.~Tamura {\it et al.}: Phys. Rev. Lett. \textbf{84} 
(2000) 5963. 

\bibitem{JLabL9Li} G.M.~Urciuoli {\it et al.} (JLab Hall A Collaboration): 
Phys. Rev. C \textbf{91} (2015) 034308. 

\bibitem{JLabL10Be} T.~Gogami {\it et al.} (JLab HKS Collaboration): 
Phys. Rev. C \textbf{93} (2016) 034314. 

\bibitem{chrien90} R.E.~Chrien {\it et al.}: Phys. Rev. C \textbf{41} 
(1990) 1062. 

\bibitem{tamura05} H.~Tamura {\it et al.}: Nucl. Phys. A \textbf{754} 
(2005) 58c. 

\bibitem{NLO13} J.~Haidenbauer, S.~Petschauer, N.~Kaiser, U.-G.~Mei{\ss}ner, 
A.~Nogga, and W.~Weise: Nucl. Phys. A \textbf{915} (2013) 24. 

\bibitem{nogga13} A.~Nogga: Nucl. Phys. A \textbf{914} (2013) 140, and 
references to earlier works cited therein. 

\bibitem{galmil13} A.~Gal and D.J.~Millener: Phys. Lett. B \textbf{725} 
(2013) 445. 

\bibitem{tamura16} H.~Tamura, private communication. 

\bibitem{agnello16} M.~Agnello, E.~Botta, T.~Bressani, S.~Bufalino, and 
A.~Feliciello: Nucl. Phys. A \textbf{954} (2016) 176. 

\end{thebibliography}
\end{document}